\documentclass[nofootinbib]{revtex4}
\usepackage[centertags]{amsmath}
\usepackage{amsthm}
\usepackage{bbm}
\usepackage{graphicx}

\DeclareMathOperator{\tr}{Tr}
\DeclareMathOperator{\str}{Str}
\DeclareMathOperator{\sdet}{Sdet}

\newcommand{\C}{\mathbbm{C}}
\newcommand{\I}{\mathcal{I}}
\newcommand{\J}{\mathcal{J}}
\newcommand{\dete}[2]{\det\left( {#1} \right)_{i,j=1,\ldots,{#2}}}
\newcommand{\abs}[1]{\mathopen| #1 \mathclose|}
\newcommand{\corep}{{\shortstack[l]
    {\tiny\{$m\times n$\} \{$p$\} \\ \tiny\{$q^T$\}}}}

\newtheorem{theorem}{Theorem}[section]
\newtheorem{conjecture}{Conjecture}[section]

\begin{document}

\title{Character expansion method for supergroups and extended
  superversions of the Leutwyler-Smilga and Berezin-Karpelevich
  integrals}

\author{C. Lehner and T. Wettig}
\affiliation{Institute for Theoretical Physics, University of
  Regensburg, 93040 Regensburg, Germany}
\author{T. Guhr}
\affiliation{Department of Physics, University of Duisburg-Essen,
  47048 Duisburg, Germany}
\author{Y. Wei}
\affiliation{School of Mathematical Sciences, University of
  Nottingham, Nottingham NG72RD, UK}

\date{May 9, 2008}

\begin{abstract}
  We introduce an extension of the character expansion method to the
  case of supergroups.  This method allows us to calculate a
  superversion of the Leutwyler-Smilga integral which, to the best of
  our knowledge, has not been calculated before.  We also use the
  method to generalize a previously calculated superversion of the
  Berezin-Karpelevich integral.  Our character expansion method should
  also allow for the calculation of other supergroup integrals.
\end{abstract}

\maketitle

\section{Introduction}

When solving models in field theory and statistical mechanics, one
often faces the challenge to integrate over continuous groups or
cosets. Examples can be found in a wide range of applications,
comprising fields as different as condensed matter
physics~\cite{Efetov} and quantum gravity~\cite{ADZ}.  There are also
prominent applications in random matrix theory~\cite{Mehta,GMW},
including statistical models in quantum chromodynamics~\cite{VW}.
From a mathematics viewpoint, group integration belongs to the field
of harmonic analysis~\cite{Helga1,Helga2}. Half a century ago,
Hua~\cite{HUA} obtained invariant measures for a large class of
spaces. Shortly thereafter, Harish-Chandra~\cite{HC} derived his
celebrated integration formula for Lie groups. The radial coordinates
on which these integrals depend parametrize the space of the Cartan
subalgebras.  In the unitary case, Harish-Chandra's result coincides
with the Itzykson-Zuber integral~\cite{IZ}. Unfortunately,
Harish-Chandra's formula cannot be applied to the majority of
integrals over orthogonal and unitary-symplectic groups that arise in
physics, because these latter integrals depend on radial coordinates
defining a space which is outside the group or its algebra,
respectively. Hence they can be interpreted as certain matrix
generalizations of Bessel functions, see a discussion in
Ref.~\cite{GK1}.  Group integrals related to but different from those
mentioned so far are of high relevance for applications in quantum
chromodynamics.  Particularly important are integrals of the
Berezin-Karpelevich~\cite{BK,GW96,JSV1} and
Leutwyler-Smilga~\cite{LSM} type.  Although in these cases the
integration is over unitary groups, Harish-Chandra's formula cannot be
used either, because the integrands are of a different form.

All these considerations carry over to integrals over supergroups.
Supermathematics~\cite{BER} --- in the present context often referred
to as supersymmetry --- was introduced to the theory of disordered
systems by Efetov~\cite{E83} and subsequently to random matrix theory
by Verbaarschot, Weidenm\"uller, and Zirnbauer~\cite{VZ,VWZ}.
Supersymmetry is nowadays an indispensable tool for many applications,
once more including those in quantum chromodynamics~\cite{Efetov,GMW,VW}.

At present, there are three different methods for exact calculations
of the integrals discussed above: (1) The diffusion equation method
was developed by 
Itzykson and Zuber~\cite{IZ} and in Refs.~\cite{GW96,JSV1} for
ordinary space. The Itzykson-Zuber integral was generalized to
supermathematics~\cite{G91} by a proper extension of the diffusion
equation method which was then further extended to work out the
supersymmetric Berezin-Karpelevich integral~\cite{GW96,JSV2}.  In
Ref.~\cite{GK3}, the diffusion equation method was generalized
beyond the unitary case to prove the supersymmetric Harish-Chandra
formula that had been conjectured in Refs.~\cite{Vera92,MZ96}.
This also provided a new proof for the ordinary Harish-Chandra
formula. (2) Balantekin introduced the character expansion method for
integrals over unitary groups in ordinary
space~\cite{Baha00,Baha01}. 
The method was further extended in  
Ref.~\cite{SWa}. Even some integrals over ordinary orthogonal
and unitary-symplectic groups could be calculated using a
generalization of the character expansion method~\cite{BaCa02}.  (3)
Shatashvili~\cite{Sha93} used an explicit parametrization of the
unitary group in terms of Gelfand-Tzetlin coordinates~\cite{GeTz50} to
calculate correlation functions in the Itzykson-Zuber model. This
method was extended to supersymmetry in Ref.~\cite{G96} and
considerably generalized to obtain recursive solutions for a wide
class of radial functions which include group integrals as special
cases~\cite{GK1,GK2}.

The main focus of the present contribution is an extension of the
character expansion method to supergroups.  This extension then allows
us to calculate a supersymmetric version of the Leutwyler-Smilga
integral.  The ordinary version of this integral yields the
finite-volume partition function of quantum chromodynamics in the
so-called epsilon-regime~\cite{Gasser:1987ah,LSM}, and its superversion has
applications in related (``partially quenched'') theories that contain
both fermionic and bosonic degrees of freedom.  We also
generalize the result previously obtained in Ref.~\cite{GW96} for a
supersymmetric Berezin-Karpelevich integral.

This article is organized as follows.  In Sec.~\ref{sec:statement},
the integrals in question are defined and the results of our
calculation are given.  A general outline of our character expansion
method is presented in Sec.~\ref{sec:sce}.  We then apply the method
to the calculation of the supersymmetric Leutwyler-Smilga and
Berezin-Karpelevich integrals in Sec.~\ref{sec4}.  We summarize our
results and give an outlook to further applications of the
supersymmetric character expansion method in Sec.~\ref{sec:summary}.
Three appendices are provided to collect various algebraic theorems, to
discuss a conjectured power series identity and its connection to
Richardson-Littlewood coefficients, and to present explicit examples
for the supersymmetric Leutwyler-Smilga integral.

\section{Statement of the integrals}
\label{sec:statement}

To define the notation to be used in the subsequent sections, we start
by stating the integrals that will be calculated using the character
expansion methods introduced in Sec.~\ref{sec:sce}.

The supersymmetric Leutwyler-Smilga integral is defined as
\begin{equation}
  \label{eq:ILS}
  \I_\text{LS} \equiv \int d\mu(U) \exp(\beta \str(AU + B U^{-1}))\:,
\end{equation}
where $U \in$ U$(m|n)$, $d\mu(U)$ is the invariant measure,
$\beta \in \C$, and $A$, $B$ are arbitrary $(m+n) \times (m+n)$
supermatrices.  We show that this integral is given by
\begin{align} 
  \label{eq:ls_result}
  \I_\text{LS} = \mathcal{C}_m \: \mathcal{C}_n \: \beta^{\frac{(m+n)-(m-n)^2}{2}}
  \:\frac{\dete{\lambda_j^{m+n-i} I_{m+n-i}(2 \beta \lambda_j)}{m+n}}
  {\Delta(\lambda^2_1,\ldots,\lambda^2_m)
    \Delta(\lambda^2_{m+1},\ldots,\lambda^2_{m+n})}\:,
\end{align}
where $\lambda^2_1,\ldots,\lambda^2_{m+n}$ are the eigenvalues of the
supermatrix $AB$ and $I_\nu$ is the modified Bessel function of the
first kind.  Furthermore, $\Delta(\lambda^2_1,\ldots,\lambda^2_m)
\equiv \prod_{1\leq i<j\leq m}(\lambda^2_i-\lambda^2_j)$ is the
Vandermonde determinant and
\begin{align}
  \label{eq:def_c}
  \mathcal{C}_n \equiv \prod_{k=1}^{n-1} k!\:.
\end{align}
Equation~\eqref{eq:ls_result} correctly reproduces the result obtained
in Ref.~\cite{SWa} for ordinary groups U$(m)$.

The second integral calculated in this work, the supersymmetric
Berezin-Karpelevich integral, is defined as
\begin{equation}
  \label{eq:IBK}
  \I_\text{BK} \equiv \int d\mu(U) \int d\mu(V) 
  \exp(\beta \str(UAVB+U^{-1}C V^{-1}D))\:,
\end{equation}
where $U,V \in$ U$(m|n)$, $\beta \in \C$, and $A,B,C,D$ are
arbitrary $(m+n) \times (m+n)$ supermatrices. We show that
\begin{equation}
  \label{eq:bk_result}
  \I_\text{BK} = \mathcal{C}_m^2 \: \mathcal{C}_n^2 \: \beta^{(m+n)-(m-n)^2} 
  \:\frac{\dete{I_0(2 \beta \lambda_i \mu_j)}{m} 
    \dete{I_0(2 \beta \lambda_{m+i} \mu_{m+j})}{n}}
  {B(\lambda^2;m,n) B(\mu^2;m,n)}\:,
\end{equation}
where $\lambda^2_1,\ldots,\lambda^2_{m+n}$ are the eigenvalues of the
supermatrix $BC$ and $\mu^2_1,\ldots,\mu^2_{m+n}$ are the eigenvalues
of the supermatrix $AD$. The Berezinian $B(\lambda^2;m,n)$ is given by
\begin{equation}
  \label{eq:berez}
  B(\lambda^2;m,n) = \frac{\Delta(\lambda^2_1,\ldots,\lambda^2_m) 
    \Delta(\lambda^2_{m+1},\ldots,\lambda^2_{m+n})}
  {\prod_{i=1}^m \prod_{j=1}^n (\lambda^2_i-\lambda^2_{m+j})}\:.
\end{equation}
This extends the result obtained in Ref.~\cite{GW96} using the
diffusion equation method. 

Note that in the above results we have to assume that the
supermatrices $AB$ (in case of $\I_\text{LS}$) or $BC$ and $AD$ (in
case of $\I_\text{BK}$) are diagonalizable in the sense explained in
App.~\ref{sec:1p1dimsm}.  However, our results can be extended to
non-diagonalizable supermatrices using a limiting procedure which is
discussed in App.~\ref{sec:1p1dimsm}.  Furthermore, the case of
coinciding eigenvalues also requires a limiting procedure, which is
discussed in App.~\ref{sec:coincev}.

We prove our results in Secs.~\ref{sec:ls} and \ref{sec:bk}.
First,
however, we discuss how the character expansion method can be extended
to supersymmetric integrals in general.

\section{Character expansion for supergroups} 
\label{sec:sce}

The concept of the character expansion method for integrals over
supergroups remains the same as for integrals over ordinary groups. We
expand the integrand in terms of supergroup characters, use
orthogonality relations of supermatrix representation elements in
order to perform the integral, and identify the remaining power series.

We will consider integrals over a supermatrix $U \in$ U$(m|n)$.
The integrands in question contain terms of the form
\begin{align}
  \exp(\beta \str(AU)) = \sum_{n=0}^\infty \frac{\beta^n}{n!} \str(AU)^n\:,
\end{align}
where $\beta \in \C$ and $A \in$ Gl$(m|n)$.

In order to expand the integrand in terms of supercharacters of $AU$
we make use of Balantekin's observation\footnote{In Ref.~\cite{B84}
  Balantekin only considered U$(m|n)$, but his arguments also
  apply to Gl$(m|n)$.} \cite{B84} that
\begin{align} 
  \label{eq:strninxi}
  \str(AU)^n = \sum_{t, \abs{t}=n} \sigma_t \, \xi_t(AU)\:,
\end{align}
where the sum is over all Young diagrams $t$ with $n$ boxes that
correspond to covariant representations\footnote{I.e., representations that are constructed only from covariant bases by symmetrization of bases according to Young diagram $t$.} and $\xi_t$ is the
corresponding supercharacter.
The expansion coefficient $\sigma_t$ of a Young diagram $t$ with $\hat{N}$ rows is given by \cite{B84}
\begin{align}
  \sigma_t \equiv \abs{t}!
  \frac{\Delta(\hat{k}_1,\dots,\hat{k}_{\hat{N}})}{\prod\limits_{i=1}^{\hat{N}}
    \hat{k}_i !}\:, 
\end{align}
where
\begin{align} 
    \hat{k}_i \equiv \hat{N}+t_i-i\:,
\end{align}
 and $t_i$ is the number of boxes in the $i$-th row of the Young
diagram. The character of a supergroup element in representation $t$ is
defined as
\begin{equation}
  \xi_t (U) \equiv \str(\Gamma^t(U)) 
  = \sum_i (-1)^{\epsilon(i)} \Gamma^t_{ii}(U)
\end{equation}
with the supermatrix representation $\Gamma^t$, $\epsilon(i)=1$ if the
index $i$ is fermionic and $\epsilon(i)=0$ otherwise.  We are thus
able to express the right-hand side of Eq.~\eqref{eq:strninxi} in
terms of the supermatrix representation elements $\Gamma^t_{ab}(U)$.

Note that our integrands always contain supertraces involving $U$ as
well as $U^{-1}$ in the exponentials. With the help of the
orthogonality relations of supermatrix representation
elements~\cite{BER},
\begin{equation}
  \label{eq:ortho}
  \int d\mu(U) \Gamma^r_{ab}(U) \Gamma^s_{cd}(U^{-1}) 
  = \alpha_{r} \delta_{rs}  \delta_{ad}  \delta_{bc} (-1)^{\epsilon(c)}\:,
\end{equation}
where $\alpha_r$ is the norm of the supergroup representation $r$, we
are thus able to perform the integral over $U$.  A careful treatment
of signs is necessary due to the anticommutation of some supermatrix
elements. For two supermatrices $A$ and $B$ this can be expressed by
the formula
\begin{equation}
  A_{ab} B_{cd} = B_{cd} A_{ab} 
  (-1)^{[\epsilon(a)+\epsilon(b)][\epsilon(c)+\epsilon(d)]}\:.
\end{equation}

We implicitly use the fact that the irreducible representations in the sum of Eq.~\eqref{eq:strninxi}
are constructed from their respective fundamental representation in the same way for U$(m|n)$ and Gl$(m|n)$ \cite{B81}.

After the integration over $U$ has been performed using
Eq.~\eqref{eq:ortho}, a power series corresponding to
Eq.~\eqref{eq:strninxi} remains.  Because of the appearance of the
factor $\alpha_r$ in Eq.~\eqref{eq:ortho}, only non-degenerate
covariant supergroup representations (i.e., those with $\alpha_r \neq
0$) contribute to this power series.  In order to identify the power
series with a known function we need explicit formulas for the
supercharacter $\xi_t$ and the norm $\alpha_t$ of a given
representation $t$.  In 1981, Balantekin and Bars obtained an integral
formula for characters of supergroups \cite{B81, B81B}. In the given
form, however, it is not suited for an application to character
expansion methods. In 1997, Alfaro, Medina, and Urrutia \cite{AMU}
obtained another formula for the supercharacters of a non-degenerate
covariant supergroup representation, which is more useful in the
present context. The restriction to non-degenerate representations
does not pose any problem for the character expansion method because,
as we have just seen, only these representations contribute to the
result.

\begin{figure}[ht]
  \includegraphics[width=5cm]{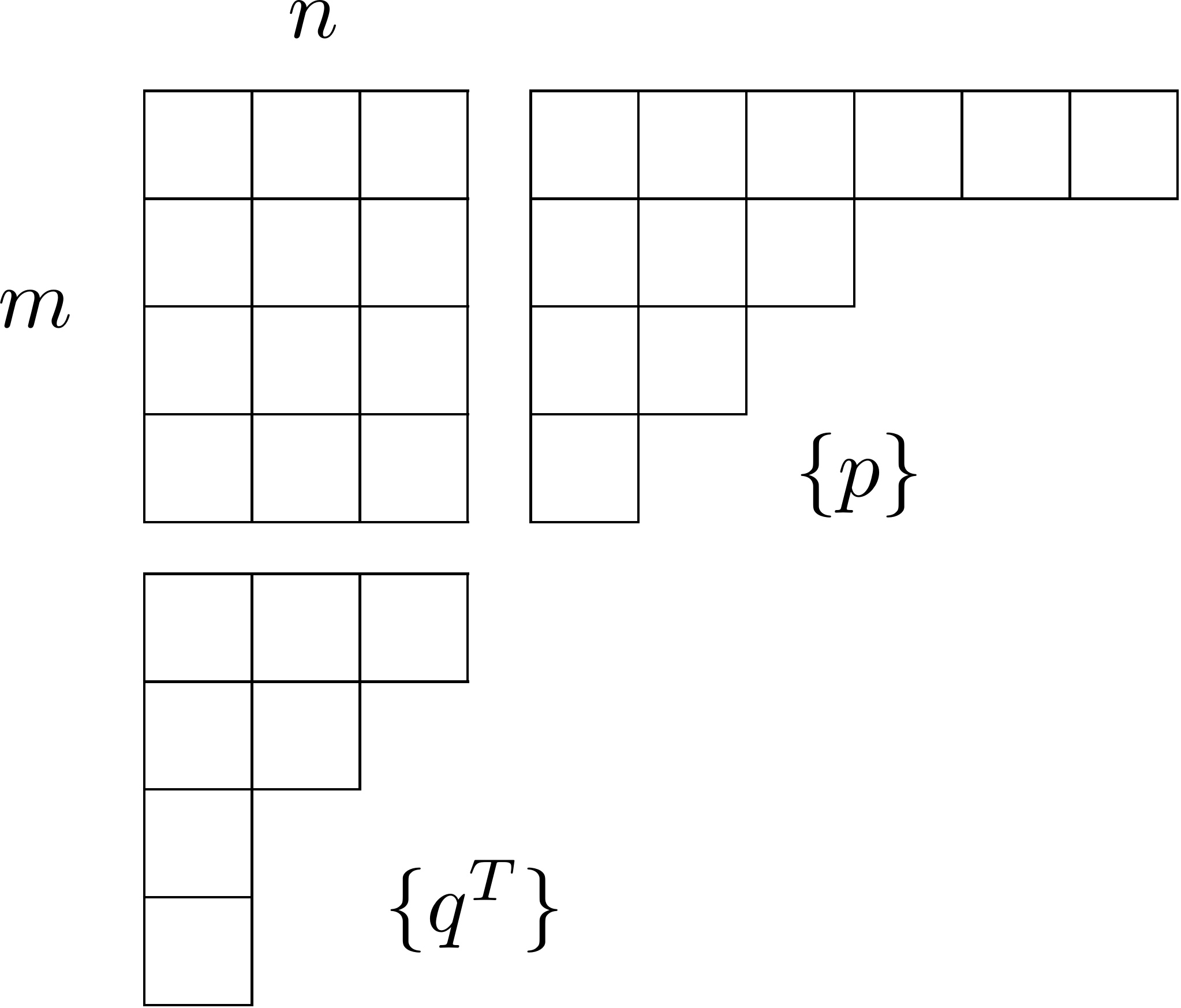}
  \caption{Young diagram corresponding to a non-degenerate covariant
    representation of Gl$(m|n)$.}
  \label{fig:nondegyd}
\end{figure}

Alfaro, Medina, and Urrutia observed that any Young diagram $t$
describing a non-degenerate covariant representation consists of an $m
\times n$ block in the top left corner and two sub-diagrams $p$ and
$q^T$ such that $p$ and $q$ are legitimate Young diagrams of Gl($m$)
and Gl($n$), respectively, cf.~Fig.~\ref{fig:nondegyd}. The sub-diagram
$p$ to the right of the $m \times n$ block is thus restricted to $m$
rows, while the sub-diagram $q$ that appears transposed below the $m
\times n$ block is restricted to $n$ rows.

The supercharacter of an $(m+n)$-dimensional supermatrix $A$ with
eigenvalues $a_1,\ldots,a_{m+n}$ corresponding to $t$ is given by
\cite{AMU}
\begin{equation} 
  \label{eq:character}
  \xi_t(A) = (-1)^{\abs{q}} \Sigma(a;m,n) \chi_p(a_1,\ldots,a_m)
  \chi_q(a_{m+1},\ldots,a_{m+n})\:,
\end{equation}
where $\abs{q}$ is the number of boxes in the sub-diagram $q$,
\begin{equation}
  \Sigma(a;m,n) \equiv \prod_{i=1}^m \prod_{j=1}^n (a_i-a_{m+j})\:,
\end{equation}
and $\chi_p$ and $\chi_q$ are the characters of the ordinary groups
Gl$(m)$ and Gl$(n)$ corresponding to the Young diagrams $p$ and
$q$. Weyl's character formula \cite{WEY} states
\begin{equation} 
  \label{eq:weyl}
  \chi_p (a_1,\ldots,a_m) 
  = \frac{\dete{a_i^{k_j}}{m}}{\Delta(a_1,\ldots,a_m)}\:,
\end{equation}
where
\begin{equation}
  k_i \equiv m+p_i-i
\end{equation}
for $1 \leq i \leq m$ and $p_i$ is the number of boxes in the $i$-th
row of the Young diagram $p$.

The norm $\alpha_t$ of the representation shown in
Fig.~\ref{fig:nondegyd} depends on the sub-diagrams in the following
manner \cite{AMU}, 
\begin{equation} 
  \label{eq:norm}
  \alpha_t = (-1)^{\abs{q}} \frac{\abs{t}!}{\abs{p}! \abs{q}!}
  \frac{\sigma_p \sigma_q}{\sigma_t} \frac{1}{d_p d_q}\:,
\end{equation}
where $\abs{p}$, $\abs{q}$, and $\abs{t}$ count the boxes in the
corresponding (sub-)diagram and $d_p$ and $d_q$ give the dimensions of
the representations of Gl$(m)$ and Gl$(n)$ corresponding to Young
diagrams $p$ and $q$, e.g.,
\begin{equation} 
  \label{eq:dim}
  d_p = \frac{\Delta(k_1,\ldots,k_m)}{\prod\limits_{i=1}^m (m-i)!}\:.
\end{equation}

We can now replace the sum over all non-degenerate, covariant
representations $t$ that remains from Eq.~\eqref{eq:strninxi} after
integration over $U$ by two sums over the sub-diagrams $p$ and $q$. An
explicit way to write these sum is, e.g.,
\begin{equation}
\sum_{k_1>k_2>\ldots>k_m\geq 0}
\end{equation}
for the sum over $p$ and
\begin{equation}
\sum_{k_{m+1}>k_{m+2}>\ldots>k_{m+n}\geq 0}
\end{equation}
for the sum over $q$, where
\begin{equation}
k_i \equiv m+n+q_{i-m}-i
\end{equation}
for $m < i \leq m+n$ and $q_{i-m}$ is the number of boxes in the
$(i-m)$-th row of the Young diagram $q$.

This concludes the extension of the character expansion method to the
case of supergroups.  We now apply these considerations to the
calculation of our supersymmetric integrals in Secs. \ref{sec:ls} and
\ref{sec:bk}.

\section{Calculation of the integrals} 
\label{sec4}

The supersymmetric extensions of the Leutwyler-Smilga and
Berezin-Karpelevich integrals are calculated in Secs.~\ref{sec:ls}
and~ \ref{sec:bk}, respectively.

\subsection{Supersymmetric Leutwyler-Smilga integral} 
\label{sec:ls}

\begin{figure}[t]
  \includegraphics[width=5cm]{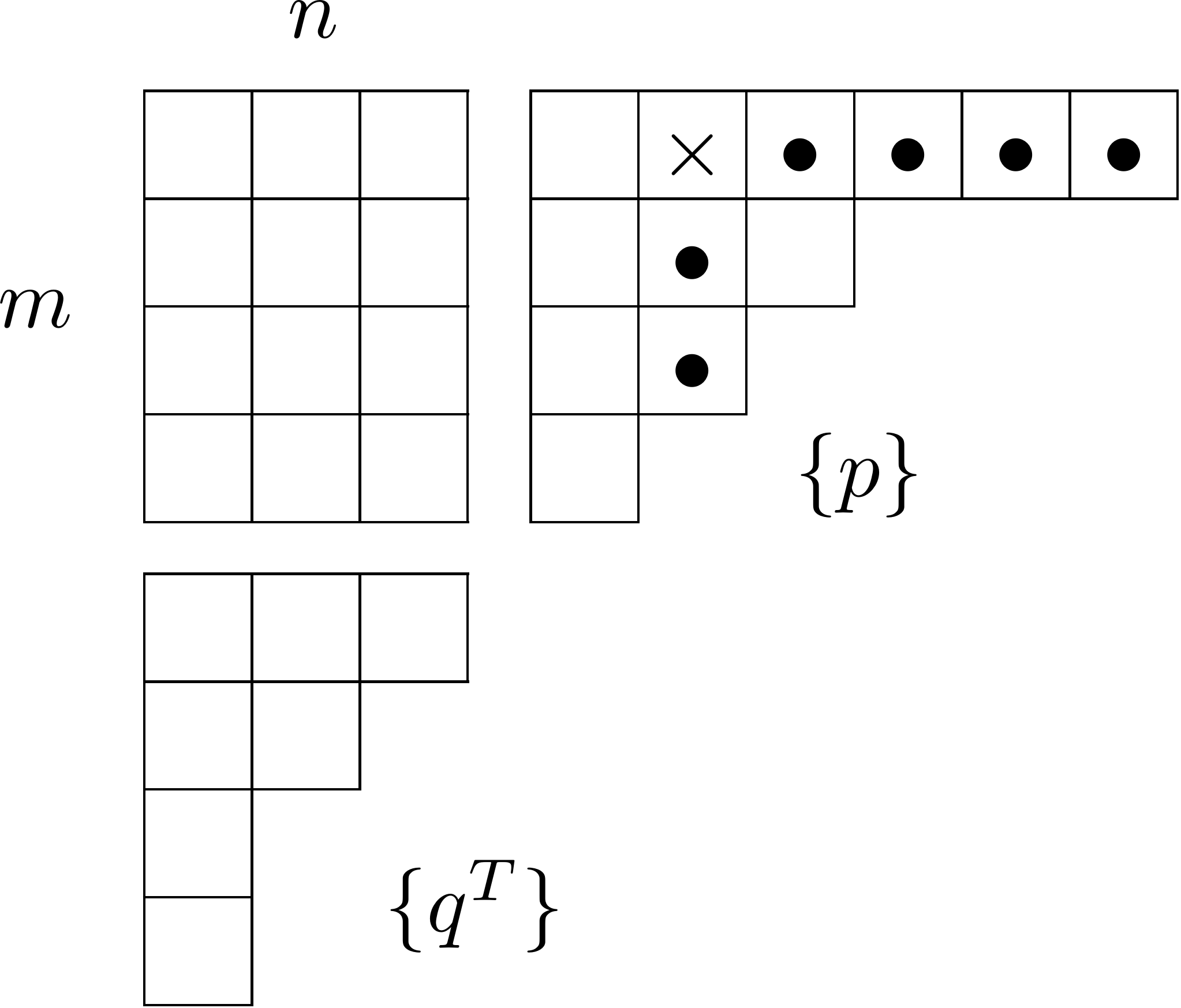}
  \caption{Young diagram for a representation $t$ with a hook of length 7.}
  \label{fig:hl}
\end{figure}

For convenience, we repeat the definition of the supersymmetric
Leutwyler-Smilga integral given in Sec.~\ref{sec:statement},
\begin{equation}
  \tag{\ref{eq:ILS}}
  \I_\text{LS} = \int d\mu(U) \exp(\beta \str(AU + B U^{-1}))\:.
\end{equation}
The notation is as in Sec.~\ref{sec:statement}.
We now apply the character expansion method laid out in
Sec.~\ref{sec:sce}. Let us expand the integrand in terms of
supercharacters and use the orthogonality of representation matrix
elements,
\begin{align}
  \I_\text{LS} &= \int d\mu(U) \exp(\beta \str(AU)) 
  \exp(\beta \str(B U^{-1})) \notag\\
  &= \int d\mu(U) \left(\sum_t \frac{\sigma_t}{\abs{t}!}
    \beta^{\abs{t}} \xi_t(AU) \right) \left(\sum_{t'}
    \frac{\sigma_{t'}}{\abs{t'}!} \beta^{\abs{t'}} \xi_{t'}(BU^{-1})
  \right) \notag\\
  &= \sum_{t,t'} \frac{\sigma_{t} \sigma_{t'}}{\abs{t}!\abs{t'}!}
  \beta^{\abs{t}+\abs{t'}} \Gamma^t_{ij}(A) \Gamma^{t'}_{lm}(B)
  (-1)^{\epsilon(i)+\epsilon(l)+(\epsilon(i)+\epsilon(j))
    (\epsilon(l)+\epsilon(m))}
  \underbrace{ \int d\mu(U) \Gamma^t_{ji}(U)
    \Gamma^{t'}_{ml}(U^{-1})}_{= \delta_{t t'}
    \delta_{jl} \delta_{im} (-1)^{\epsilon(m)} \alpha_t} \notag\\
  &=\sum_t \left( \frac{\sigma_t}{\abs{t}!}
    \beta^{\abs{t}} \right)^2 \alpha_t \xi_t(AB)\:,
\end{align}
where the sums are over all covariant representations $t$, $t'$ and
$\Gamma^t(U)$ is the representation matrix of $U$ corresponding to the
Young diagram $t$. As already pointed out in Sec.~\ref{sec:sce}, we
can restrict the sum over the covariant representations $t$ to the
non-degenerate ones, for which $\alpha_t\neq 0$. The sum over all
non-degenerate representations can be expressed by two sums over all
sub-diagrams $p$ and $q$ corresponding to the ordinary groups Gl$(m)$
and Gl$(n)$, respectively. Inserting the norm given in
Eq.~\eqref{eq:norm} and the supercharacter given in
Eq.~\eqref{eq:character} yields
\begin{align}
  \I_\text{LS} &= \mathcal{C}_m \: \mathcal{C}_n \: \sum_{p,q}
  \frac{1}{(\abs{p}+\abs{q}+mn)!} \sigma_\corep 
  \left(\prod_{i=1}^{m+n} \frac{1}{k_i!} \right) \Sigma(\lambda^2;m,n)
  \chi_p(\lambda^2_1,\ldots,\lambda^2_m) 
  \chi_q(\lambda^2_{m+1},\ldots,\lambda^2_{m+n})\:,
\end{align}
where $\lambda^2_1,\ldots,\lambda^2_{m+n}$ are the eigenvalues of the
supermatrix $AB$ and
\begin{equation}
  \frac{\sigma_p}{\abs{p}! d_p} \frac{\sigma_q}{\abs{q}! d_q}  =
  \mathcal{C}_m \: \mathcal{C}_n \: \left(\prod_{i=1}^{m+n} \frac{1}{k_i!} \right)\:.
\end{equation}
Note that we have set $\beta=1$, as we can reinstate it later by a
redefinition of $A$ and $B$. The term
\begin{equation}
 \frac{\sigma_t}{\abs{t}!} = \frac{1}{(\abs{p}+\abs{q}+mn)!} \sigma_\corep
\end{equation}
can be decomposed in terms of the sub-diagrams $p$ and $q$ using the
hook length formula
\begin{equation} 
  \label{eq:hl}
  \frac{\sigma_t}{\abs{t}!} =
  \prod_{ij}\frac{1}{h_{ij}}\:,
\end{equation}
where the product is over all boxes in the diagram $t$.  This formula
can be found in standard textbooks. The hook length $h_{ij}$ of the
$j$-th box in the $i$-th line of the diagram is defined as the number
of boxes to the right of this box plus the number of boxes below this
box plus one, cf.~Fig.~\ref{fig:hl}.
The decomposed $\sigma_t$ reads
\begin{align}
  \frac{\sigma_t}{\abs{t}!} &= \frac{\sigma_p}{\abs{p}!}
  \frac{\sigma_q}{\abs{q}!} \left( \prod_{i=1}^m \prod_{j=1}^n
    \frac{1}{m+n+1+p_i-i+q_j-j} \right) \notag\\ 
  &= \frac{\sigma_p}{\abs{p}!} \frac{\sigma_q}{\abs{q}!} \left(
    \prod_{i=1}^m \prod_{j=1}^n \frac{1}{k_i + k_{m+j} +1} \right)\:.
\end{align}
Putting the pieces together we obtain
\begin{align} 
  \I_\text{LS}  &= \mathcal{C}_m \: \mathcal{C}_n \: \sum_{p,q}
  \frac{\Delta(k_1,\ldots,k_m)\Delta(k_{m+1},\ldots,k_{m+n})}
  {(k_1!)^2 \cdots (k_{m+n}!)^2 } 
  \left( \prod_{i=1}^m \prod_{j=1}^n \frac{1}{k_i +
      k_{m+j} +1} \right)  \notag\\ 
  \label{eq:hookconj}
  &\quad  \times \Sigma(\lambda^2;m,n) \chi_p(\lambda^2_1,\ldots,\lambda^2_m)
  \chi_q(\lambda^2_{m+1},\ldots,\lambda^2_{m+n})\:.
\end{align}
Using Weyl's character formula, given in Eq.~\eqref{eq:weyl}, we write
the sums over $p$ and $q$ as sums over $k_1,\ldots,k_{m+n}$.  We then
apply Theorem~\ref{th:rearrangement} and are left with
\begin{align}
  \I_\text{LS}  &=\mathcal{C}_m \: \mathcal{C}_n \:
  \frac{\Sigma(\lambda^2;m,n)}{\Delta(\lambda^2_1,\ldots,\lambda^2_m)
    \Delta(\lambda^2_{m+1},\ldots,\lambda^2_{m+n})}  \notag\\
  &\quad \times \sum_{k_1,\ldots,k_{m+n}=0}^\infty
  \frac{\Delta(k_1,\ldots,k_m)\Delta(k_{m+1},\ldots,k_{m+n})}{(k_1!)^2
    \cdots (k_{m+n}!)^2} 
  \left( \prod_{i=1}^m \prod_{j=1}^n \frac{1}{k_i + k_{m+j} +1} \right)
  \lambda_1^{2 k_1} \cdots \lambda_{m+n}^{2 k_{m+n}}\:.
\end{align}

We now use a power series identity which is conjectured in
App.~\ref{sec:dps}. It states that the two power series $\J_0$ and
$\J_m$ of $N$ complex variables $z_1,\ldots,z_N$, defined by
\begin{equation}
  \label{eq:lspdef}
  \J_0 \equiv \sum_{k'_1,\ldots,k'_N=0}^\infty
  \frac{\Delta(k'_1,\ldots,k'_N)}{(k'_1!)^2 \cdots (k'_N!)^2}
  \: z_1^{k'_1} \cdots z_N^{k'_N}
\end{equation}
and
\begin{equation}
  \label{eq:Jm}
  \J_m \equiv \sum_{k_1,\ldots,k_N=0}^\infty
  \frac{\Delta(k_1,\ldots,k_m) \Delta(k_{m+1},\ldots,k_N)}{(k_1!)^2
    \cdots (k_N!)^2} \left( \prod_{i=1}^m \prod_{j=m+1}^N
    \frac{z_i-z_j}{k_i+k_j+1} \right) z_1^{k_1} \cdots z_N^{k_N}\:,
\end{equation}
are identical for all $m$ with $1\le m\le N$.  Unfortunately, we do not
have an analytical proof for this identity for arbitrary $N$.
Explicit proofs are given for $N=2$ and $N=3$ in Apps.~\ref{sec:dps}
and \ref{sec:lsexpl}.  In App.~\ref{sec:dps} we present compelling
numerical and partial analytical evidence in favor of the conjecture
for arbitrary $N$, which leaves very little doubt that the conjecture
is correct.  Appendix~\ref{sec:dps} also discusses a connection of the
identity to Richardson-Littlewood coefficients.

Inserting the conjectured identity $\J_0=\J_m$ in
Eq.~\eqref{eq:hookconj}, we obtain
\begin{align}
  \I_\text{LS} &= \mathcal{C}_m \: \mathcal{C}_n \: 
  \frac{1}{\Delta(\lambda^2_1,\ldots,\lambda^2_m)
    \Delta(\lambda^2_{m+1},\ldots,\lambda^2_{m+n})}  
  \sum_{k_1,\ldots,k_{m+n}=0}^\infty 
  \frac{\Delta(k_1,\ldots,k_{m+n})}{(k_1!)^2 \cdots (k_{m+n}!)^2} 
  \lambda_1^{2 k_1} \cdots \lambda_{m+n}^{2 k_{m+n}}\:.
\end{align}
Applying Theorems~\ref{th:rearrangement} and \ref{th:feit} then yields
\begin{align} 
  \I_\text{LS} &= \mathcal{C}_m \: \mathcal{C}_n \:
  \frac{1}{\Delta(\lambda^2_1,\ldots,\lambda^2_m)
    \Delta(\lambda^2_{m+1},\ldots,\lambda^2_{m+n})} \notag \\
  &\quad \times
  \sum_{k_1>k_2>\dots>k_{m+n} \geq 0} \dete{\frac{1}{k_j! (k_j-m-n+i)!}}{{m+n}}
  \dete{\lambda_i^{2 k_j}}{{m+n}}\:.
\end{align}
Using Theorem~\ref{th:powerseriesa} and the power series expansion of
the Bessel function \cite{ABR},
\begin{equation}
  \lambda^{-\nu} I_\nu (2 \lambda)  =
  \sum_{k=0}^\infty \frac{\lambda^{2k}}{k! (k+\nu)!}\:,
\end{equation}
we find
\begin{align} 
  \I_\text{LS} &= \mathcal{C}_m \: \mathcal{C}_n \:
  \frac{\dete{\lambda_j^{{m+n}-i} I_{i-m-n}(2 \lambda_j)}{{m+n}}}
  {\Delta(\lambda^2_1,\ldots,\lambda^2_m)
    \Delta(\lambda^2_{m+1},\ldots,\lambda^2_{m+n})}\notag\\
  &= \mathcal{C}_m \: \mathcal{C}_n \:
  \frac{\dete{\lambda_j^{{m+n}-i} I_{{m+n}-i}(2 \lambda_j)}{{m+n}}}
  {\Delta(\lambda^2_1,\ldots,\lambda^2_m)
    \Delta(\lambda^2_{m+1},\ldots,\lambda^2_{m+n})}\:,
\end{align}
where we used $I_\nu = I_{-\nu}$ in the last line. Finally, we
reinstate $\beta$ by rescaling the $\lambda_i$ and obtain
\begin{align}
  \tag{\ref{eq:ls_result}}
  \I_\text{LS} &= \mathcal{C}_m \: \mathcal{C}_n \:
  \beta^{\frac{(m+n)-(m-n)^2}{2}}\:
  \frac{\dete{\lambda_j^{{m+n}-i} I_{{m+n}-i}(2 \beta \lambda_j)}{{m+n}}}
  {\Delta(\lambda^2_1,\ldots,\lambda^2_m)
    \Delta(\lambda^2_{m+1},\ldots,\lambda^2_{m+n})} \:.
\end{align}
To the best of our knowledge this result has not been derived
before.

\subsection{Extended Supersymmetric Berezin-Karpelevich Integral} 
\label{sec:bk}

We now calculate the extension of the Berezin-Karpelevich integral
over $U,V \in$ U$(m|n)$ defined in Sec.~\ref{sec:statement},
\begin{equation}
  \tag{\ref{eq:IBK}}
  \I_\text{BK} \equiv \int d\mu(U) \int d\mu(V) 
  \exp(\beta \str(UAVB+U^{-1}C V^{-1}D))\:,
\end{equation}
where again the notation is as in Sec.~\ref{sec:statement}. We expand
the integrand in terms of supercharacters and obtain
\begin{equation}
  \I_\text{BK} = \sum_{t,t'} \frac{\sigma_t \sigma_{t'}}{\abs{t}! \abs{t'}!}
  \beta^{\abs{t}+\abs{t'}} \int d\mu(U) \int d\mu(V) \: \xi_t(UAVB)
  \xi_{t'}(U^{-1}C V^{-1} D)\:,
\end{equation}
where the sum is over all covariant representations of Gl$(m|n)$.  We
now insert the definition of the supercharacter as the supertrace of
the corresponding representation matrix and make use of the
orthogonality of representation matrix elements. After a careful
handling of signs due to supermatrix element commutation, we obtain
\begin{equation}
  \I_\text{BK} = \sum_{t} 
  \left(\frac{\sigma_t \alpha_t}{\abs{t}!} \beta^{\abs{t}}\right)^2
  \xi_t(BC) \xi_t(AD)\:. 
\end{equation}
We again replace the sum over all covariant representations $t$ by two
sums over all representations $p$ and $q$ of the ordinary groups
Gl$(m)$ and Gl$(n)$, respectively, as the terms we omit in doing so are
terms with $\alpha_t=0$, which do not contribute to the
result. Inserting Eqs.~\eqref{eq:norm} and \eqref{eq:character}, we
obtain
\begin{align} 
  \I_\text{BK} &= 
  \Sigma(\lambda^2;m,n) \Sigma(\mu^2;m,n) \sum_{p,q} 
  \left(\frac{\beta^{\abs{q}+\abs{p}+mn}}{\abs{q}!\abs{p}!}
    \frac{\sigma_p \sigma_q}{ d_p d_q} \right)^2 \notag\\
  &\quad \times \chi_p(\lambda^2_1,\ldots,\lambda^2_m)
  \chi_p(\mu^2_1,\ldots,\mu^2_m)
  \chi_q(\lambda^2_{m+1},\ldots,\lambda^2_{m+n}) 
  \chi_q(\mu^2_{m+1},\ldots,\mu^2_{m+n}) \notag\\
  &= 
  \Sigma(\lambda^2;m,n) \Sigma(\mu^2;m,n) \beta^{2mn} 
  \left[\sum_p \frac{\beta^{2 \abs{p}} \sigma_p^2}{d_p^2 (\abs{p}!)^2}
    \chi_p(\lambda^2_1,\ldots,\lambda^2_m)
    \chi_p(\mu^2_1,\ldots,\mu^2_m) \right] \notag\\
  &\quad \times \left[\sum_q \frac{\beta^{2 \abs{q}} \sigma_q^2}{d_q^2
      (\abs{q}!)^2}\chi_q(\lambda^2_{m+1},\ldots,\lambda^2_{m+n})
    \chi_q(\mu^2_{m+1},\ldots,\mu^2_{m+n}) \right]\:,
\end{align}
where $\lambda^2_1,\ldots,\lambda^2_{m+n}$ are the eigenvalues of the
matrix $BC$ and $\mu^2_1,\ldots,\mu^2_{m+n}$ are the eigenvalues of
the matrix $AD$, respectively. According to Eq.~(21) of
Ref.~\cite{SWa}, we find that the terms in brackets are nothing but
ordinary Berezin-Karpelevich integrals over groups U$(m)$ and U$(n)$,
for which the result is given in Ref.~\cite{SWa}. We thus obtain
\begin{equation}
  \tag{\ref{eq:bk_result}}
  \I_\text{BK} =\mathcal{C}_m^2 \: \mathcal{C}_n^2 \:  \beta^{(m+n)-(m-n)^2}\:
  \frac{\dete{I_0(2\beta \lambda_i \mu_j)}{m} \dete{I_0(2 \beta 
      \lambda_{m+i} \mu_{m+j})}{n}}{B(\lambda^2;m,n) B(\mu^2;m,n)}
\end{equation}
with the Berezinian 
\begin{equation}
  \tag{\ref{eq:berez}}
  B(\lambda^2;m,n) = \frac{\Delta(\lambda^2_1,\ldots,\lambda^2_m)
    \Delta(\lambda^2_{m+1},\ldots,\lambda^2_{m+n})}
  {\prod_{i=1}^m \prod_{j=1}^n (\lambda^2_i-\lambda^2_{m+j})}\:.
\end{equation}
This extends the result obtained in Ref.~\cite{GW96} using the
diffusion equation method by introducing two additional independent
supermatrix parameters.

\section{Summary and outlook} 
\label{sec:summary} 

We have extended the character expansion method to supergroup
integrals and calculated a supersymmetric Leutwyler-Smilga integral
that, to the best of our knowledge, has not been known before. In the
course of the calculation of this integral, we used a conjecture
which, unfortunately, lacks a complete proof.  However, in
App.~\ref{sec:dps} we presented strong arguments that this conjecture
is correct.  We also calculated a supersymmetric Berezin-Karpelevich
integral extending the result obtained in Ref.~\cite{GW96}.

The character expansion method developed in this paper should allow
for the calculation of other supergroup integrals that might not be
calculable using other methods such as the diffusion equation method
\cite{GW96}. In the case of the ordinary Leutwyler-Smilga integral, a
determinant term can be included in the integrand as well
\cite{JSV1,Akuzawa:1998tq,SWa}.  The inclusion of a superdeterminant
term is also of interest in the supersymmetric version of the
integral. Work in this direction is in progress.

\begin{acknowledgments}
  CL would like to thank Andreas Sch\"afer, who suggested the main
  idea of App.~\ref{sec:partialproof}.  TG acknowledges support from
  Deutsche Forschungsgemeinschaft (Sonderforschungsbereich Transregio
  12, ``Symmetries and Universality in Mesoscopic Systems'').  
  YW acknowledges support from EPSRC grant EP/C515056/1 
  (``Random Matrices and Polynomials: a tool to understand complexity'').
\end{acknowledgments}

\appendix

\section{Extension to non-diagonalizable supermatrices}
\label{sec:1p1dimsm}
The integral formulas presented in Sec.~\ref{sec:statement} can be extended to non-diagonalizable supermatrices by a limiting procedure.  In this section we demonstrate the problem of non-diagonalizable supermatrices in the case of $1+1$ dimensions.  The results can be generalized in a straightforward way to higher dimensions.

Let $M$ be an arbitrary $(1+1)$-dimensional supermatrix,
\begin{align}
  M =
  \begin{pmatrix}
    a & \alpha \\ \beta & b
  \end{pmatrix}\:,
\end{align}
where $a$, $b$ are even elements of the algebra and $\alpha$, $\beta$ are odd
elements of the algebra. If $(a-b)^{ -1}$ exists, i.e., $a$ and $b$
are not equal up to nilpotent terms, $M$ can be diagonalized,
\begin{align}
  M \equiv V_M^{-1} M_D V_M\:,
\end{align}
where
\begin{align}
  V_M & =
  \begin{pmatrix}
    1 - \frac{\alpha \beta}{2 (a-b)^2} & -\frac{\alpha}{a-b} \\ \frac{\beta}{a-b} & 1 + \frac{\alpha \beta}{2 (a-b)^2}
  \end{pmatrix}\:,\\
V_M^{-1} & = 
\begin{pmatrix}
    1 - \frac{\alpha \beta}{2 (a-b)^2} & \frac{\alpha}{a-b} \\ -\frac{\beta}{a-b} & 1 + \frac{\alpha \beta}{2 (a-b)^2}
\end{pmatrix}\:,\\
M_D & =
\begin{pmatrix}
 a + \frac{\alpha \beta}{a-b} & 0 \\
0 & b + \frac{\alpha \beta}{a-b}
\end{pmatrix}
\:.
\end{align}
The entries of $M_D$ are the (geometric) eigenvalues of $M$.

We would like to briefly comment on the algebraic definition of eigenvalues as solutions of the characteristic equations $\sdet(M-m) = 0$ and $\sdet(M-m)^{-1} = 0$. In the present case the algebraic definition has four solutions,
\begin{align}
  m_1& = a + \frac{\alpha \beta}{a-b}\:, \\
  m_2& = b + \frac{\alpha \beta}{a-b}\:, \\
  m_3& = b - \frac{\alpha \beta}{a-b}\:, \\
  m_4& = a - \frac{\alpha \beta}{a-b}\:.
\end{align}
Only two of these coincide with the geometric eigenvalues.  Therefore a consistent definition of eigenvalues is only possible if a supermatrix is diagonalizable.  We assume the geometric definition of eigenvalues in this paper.
In general a supermatrix cannot be diagonalized if one of the eigenvalues of its boson-boson block coincides with one of the eigenvalues of its fermion-fermion block up to nilpotent terms.
This problem looks severe at first sight, but although the limit $b \to a$ of the eigenvalues does not exist,
the supersymmetric integrals have well-defined limits.

Let us consider the case of $b = a + \varepsilon$ with $\varepsilon \to 0$. The case of $b \to a + \hat{n}$, where $\hat{n}$ is nilpotent, can be obtained in the same way by keeping higher orders in $\varepsilon$ and substituting $\hat{n}$ for $\varepsilon$ in the end.

In the case of the supersymmetric Leutwyler-Smilga integral given in Eq.~\eqref{eq:ls_result} with $AB=M$, we find
\begin{align}
 I_0(\lambda_1) I_1(\lambda_2) \lambda_2 -
 I_0(\lambda_2) I_1(\lambda_1) \lambda_1 \to
 - \frac{\alpha \beta}{2 a} I_1(\sqrt{a}))^2\:.
\end{align}

The supersymmetric Berezin-Karpelevich integral given in Eq.~\eqref{eq:bk_result} with $BC=M$ also has a well-defined limit for $b \to a$ with
\begin{align}
  (\lambda_1^2 - \lambda_2^2) I_0(\lambda_1 \mu_1) I_0(\lambda_2 \mu_2) \to
\frac{\alpha \beta}{2 \sqrt{a}} \left( I_0(\sqrt{a} \mu_2) I_1(\sqrt{a} \mu_1) \mu_1
 + I_0(\sqrt{a} \mu_1) I_1(\sqrt{a} \mu_2) \mu_2 \right).
\end{align}

As mentioned above, this limiting procedure can be generalized to
higher dimensions.

\section{The limit of coinciding eigenvalues}
\label{sec:coincev}
The results given in Eqs.~\eqref{eq:ls_result} and
\eqref{eq:bk_result} require a limiting procedure if two eigenvalues
within the first $m$ or within the last $n$ eigenvalues coincide.  We
calculate these limits explicitly in this section.

We first consider the result of the supersymmetric Leutwyler-Smilga integral,
\begin{align}
  \tag{\ref{eq:ls_result}}
  \I_\text{LS} &= \mathcal{C}_m \: \mathcal{C}_n \:
  \beta^{\frac{(m+n)-(m-n)^2}{2}}\:
  \frac{\dete{\lambda_j^{{m+n}-i} I_{{m+n}-i}(2 \beta \lambda_j)}{{m+n}}}
  {\Delta(\lambda^2_1,\ldots,\lambda^2_m)
    \Delta(\lambda^2_{m+1},\ldots,\lambda^2_{m+n})} \:.
\end{align}
If one of the bosonic eigenvalues ($\lambda_1,\ldots,\lambda_m$) coincides with one of the fermionic eigenvalues ($\lambda_{m+1},\ldots,\lambda_{m+n}$), the integral vanishes.  The case of two coinciding eigenvalues within bosonic or fermionic eigenvalues is slightly more involved.  Let us assume, without loss of generality, that the two coinciding eigenvalues are $\lambda_1$ and $\lambda_2$.  We analyze the limit $\lambda_2 - \lambda_1 \equiv \varepsilon \to 0$.  Up to linear order in $\varepsilon$, we find
\begin{align} \notag
\lambda_2^{{m+n}-i} I_{{m+n}-i}(2 \beta \lambda_2) & =
 (\lambda_1+\varepsilon)^{{m+n}-i} I_{{m+n}-i}(2 \beta (\lambda_1+\varepsilon)) \\ &= 
\lambda_1^{{m+n}-i} I_{{m+n}-i}(2 \beta \lambda_1) + 2 \beta \varepsilon \lambda_1^{{m+n}-i} I_{{m+n}-i-1}(2 \beta \lambda_1)
\end{align}
and
\begin{align}
  \Delta(\lambda^2_1,\ldots,\lambda^2_m) = -2 \varepsilon \lambda_1  \Delta(\lambda^2_2,\ldots,\lambda^2_m) \prod_{i=3}^m (\lambda_1^2- \lambda_i^2)\:.
\end{align}
Therefore the result for $\lambda_2 = \lambda_1$ is obtained by replacing the second column of the matrix in the numerator by
\begin{align}
  \beta \lambda_1^{{m+n}-i-1} I_{{m+n}-i-1}(2 \beta \lambda_1)
\end{align}
and the first Vandermonde determinant in the denominator by
\begin{align}
  \label{eq:van}
  - \Delta(\lambda^2_2,\ldots,\lambda^2_m) \prod_{i=3}^m (\lambda_1^2- \lambda_i^2)\:.
\end{align}

The supersymmetric Berezin-Karpelevich integral 
\begin{equation}
  \tag{\ref{eq:bk_result}}
  \I_\text{BK} = \mathcal{C}_m^2 \: \mathcal{C}_n^2 \: \beta^{(m+n)-(m-n)^2}\:
  \frac{\dete{I_0(2\beta \lambda_i \mu_j)}{m} \dete{I_0(2 \beta 
      \lambda_{m+i} \mu_{m+j})}{n}}{B(\lambda^2;m,n) B(\mu^2;m,n)}
\end{equation}
also vanishes if one of the bosonic eigenvalues
($\lambda_1,\ldots,\lambda_m$ resp. $\mu_1,\ldots,\mu_m$) coincides
with one of the corresponding fermionic eigenvalues
($\lambda_{m+1},\ldots,\lambda_{m+n}$
resp. $\mu_{m+1},\ldots,\mu_{m+n}$).  For the case of two coinciding
eigenvalues within bosonic or fermionic eigenvalues, we again
consider $\lambda_2 = \lambda_1 + \varepsilon$ with $\varepsilon \to
0$.  Up to linear order in $\varepsilon$,
\begin{align}
  I_0(2 \beta \lambda_2 \mu_j) = I_0(2 \beta \lambda_1 \mu_j) + 2\beta\mu_j \varepsilon I_1(2 \beta \lambda_1 \mu_j)\:.
\end{align}
For $\lambda_2 = \lambda_1$ the result is therefore obtained by replacing the second row of the first matrix in the numerator by
\begin{align}
  \beta \mu_j \lambda_1^{-1} I_{-1}(2 \beta \lambda_1 \mu_j)
\end{align}
and the first Vandermonde determinant in the first Berezinian in the
denominator by Eq.~\eqref{eq:van}.

This procedure can be repeated in the case of multiple coinciding eigenvalues in a straightforward way.

\section{Algebraic theorems}

In the following theorems, $p_1,\ldots,p_N$ is a permutation of
$1,\ldots,N$ and $\sigma_{\lbrace p \rbrace}$ is the sign of the
permutation (positive if an even number of neighbors were exchanged,
negative otherwise).
\begin{theorem}
  \label{th:rearrangement}
  If $A_{k_1,\ldots,k_N}=\sigma_{\lbrace p \rbrace}
  A_{k_{p_1},\ldots,k_{p_N}}$, then
  \begin{equation}
    \sum_{k_1,\ldots,k_N=0}^\infty A_{k_1,\ldots,k_N} a_1^{k_1}
    \cdots a_N^{k_N} = \sum_{k_1>k_2>\dots>k_N \geq 0}
    A_{k_1,\ldots,k_N} \dete{a_i^{k_j}}{N}\:.
  \end{equation}
  \begin{proof}
    \begin{align} 
      \sum_{k_1,\ldots,k_N=0}^\infty A_{k_1,\ldots,k_N} a_1^{k_1} 
      \cdots a_N^{k_N} &= \frac{1}{N!} 
      \sum_{\lbrace p \rbrace} \sum_{k_1,\ldots,k_N=0}^\infty
      \underbrace{A_{k_{p_1},\ldots,k_{p_N}}}_{=\sigma_{\lbrace p \rbrace}
        A_{k_1,\ldots,k_N}} a_1^{k_{p_1}} \cdots a_N^{k_{p_N}} \notag\\
      & = \frac{1}{N!} \sum_{k_1,\ldots,k_N=0}^\infty  A_{k_1,\ldots,k_N}
      \underbrace{ \sum_{\lbrace p \rbrace} \sigma_{\lbrace p \rbrace}
        a_1^{k_{p_1}} \cdots a_N^{k_{p_N}}}_{=\dete{a_i^{k_j}}{N}}\:.
    \end{align}
    Now the summand is invariant under permutation of the $k_i$, so
    instead of averaging over all permutations, we can pick one
    specific order. Noting that the summand vanishes for a pair of
    equal $k$, the proof is complete.
  \end{proof}
\end{theorem}

\noindent The following theorem was taken from Ref.~\cite{HUA}.
\begin{theorem} 
  \label{th:powerseriesa}
  Let the power series
  \begin{equation}
    f_i (z) = \sum_{k=0}^\infty a_k^i z^k
  \end{equation}
  converge for $\abs{z}<\rho$. Then,
  \begin{equation}
    \dete{f_i(z_j)}{N} = \sum_{k_1>k_2>\dots>k_N \geq 0}
    \dete{a^i_{k_j}}{N} \dete{z_i^{k_j}}{N}
  \end{equation}
  for $\abs{z_i}<\rho$, $i=1,\ldots,N$.
  \begin{proof}
    \begin{align} 
      \dete{f_i(z_j)}{N} & =\sum_{\lbrace p \rbrace} \sigma_{\lbrace p \rbrace}
      f_{p_1}(z_1)\cdots f_{p_N}(z_N) \notag\\
      & = \sum_{\lbrace p \rbrace} \sigma_{\lbrace p \rbrace}
      \sum_{k_1,\ldots,k_N=0}^\infty a_{k_1}^{p_1} \cdots 
      a_{k_N}^{p_N} z_1^{k_1} \cdots z_N^{k_N} \notag\\
      & = \sum_{k_1,\ldots,k_N=0}^\infty \underbrace{\left(
          \sum_{\lbrace p \rbrace} \sigma_{\lbrace p \rbrace} 
          a_{k_1}^{p_1} \cdots a_{k_N}^{p_N} \right)}_{=
        \dete{a^i_{k_j}}{N}} z_1^{k_1} \cdots z_N^{k_N}\:.
    \end{align}
    We can now apply Theorem~\ref{th:rearrangement}, and the proof is
    complete.
  \end{proof}
\end{theorem}

\begin{theorem} 
  \label{th:feit}
  Let the integers $k_1,\ldots,k_N \geq0$ and $n_1,\ldots,n_N \geq 0$
  be related through
  \begin{equation}
    k_j \equiv  n_j + N - j \:.
  \end{equation}
  Then,
  \begin{equation}
    \dete{\frac{1}{(n_j+i-j)!}}{N} =
    \frac{\Delta(k_1,\dots,k_N)}{\prod_{i=1}^N k_i !}\:,
  \end{equation}
  where
  \begin{equation}
    \Delta(k_1,\ldots,k_N) \equiv \prod_{1\leq i < j \leq N} (k_i-k_j)
    = \dete{k_i^{N-j}}{N}\:.
  \end{equation}
  Note that terms with factorials of negative integers in the
  denominator have to be understood as equal to zero.
  \begin{proof}
    By rearranging the columns of the matrices in question it is
    straightforward to see that the statement is equivalent to
    \begin{equation}
      \dete{\frac{k_j!}{(k_j-i+1)!}}{N} = \dete{k_i^{j-1}}{N}\:.
    \end{equation}
    Let us define two matrices $A$ and $C$ with elements
    \begin{equation} 
      \begin{split}
        A_{ij} &\equiv k_j^{i-1}\:, \\
        C_{ij} &\equiv \frac{k_j!}{(k_j-i+1)!}\:.
      \end{split}
    \end{equation}
    Because of the structure of $C_{ij}$ it is always possible to find
    a triangular matrix $B$ with diagonal entries $1,\ldots,1$ that
    satisfies
    \begin{equation}
      C = B A \:.
    \end{equation}
    We thus obtain
    \begin{equation}
      \det(C) = \underbrace{\det(B)}_{=1} \det(A) = \det(A)\:,
    \end{equation}
    which completes the proof.
  \end{proof}
\end{theorem}

\section{Conjecture of a power series identity} 
\label{sec:dps}

\subsection{Statement of the conjecture and discussion}

\begin{conjecture} 
  \label{th:ls}
  Let two power series $\J_0$ and $\J_m$ of $N$ complex variables
  $z_1,\ldots,z_N$ be defined by
  \begin{equation} 
    \tag{\ref{eq:lspdef}}
    \J_0 \equiv \sum_{k'_1,\ldots,k'_N=0}^\infty
    \frac{\Delta(k'_1,\ldots,k'_N)}{(k'_1!)^2 \cdots (k'_N!)^2}
    z_1^{k'_1} \cdots z_N^{k'_N}
  \end{equation}
  and
  \begin{equation}
    \tag{\ref{eq:Jm}}
    \J_m \equiv \sum_{k_1,\ldots,k_N=0}^\infty
    \frac{\Delta(k_1,\ldots,k_m) \Delta(k_{m+1},\ldots,k_N)}{(k_1!)^2
      \cdots (k_N!)^2} \left( \prod_{i=1}^m \prod_{j=m+1}^N
      \frac{z_i-z_j}{k_i+k_j+1} \right) z_1^{k_1} \cdots z_N^{k_N}\:,
  \end{equation}
  where
  \begin{equation}
    \Delta(a_1,\ldots,a_N) \equiv \prod_{1\leq i < j \leq N} (a_i-a_j)
  \end{equation}
  is the Vandermonde determinant. Then,
  \begin{equation}
    \J_0 = \J_m \quad \text{for all} \quad 1 \le m \le N \:.
  \end{equation}
\end{conjecture}
In the case of $N=2$ and $m=1$, the proof is straightforward,
\begin{align}
  \sum_{k_1,k_2=0}^\infty \frac{k_1 - k_2}{(k_1!)^2 (k_2!)^2}
  z_1^{k_1} z_2^{k_2} &= \sum_{k_1,k_2=0}^\infty
  \frac{k_1^2 - k_2^2}{(k_1!)^2 (k_2!)^2} \frac{1}{k_1 + k_2}
  z_1^{k_1} z_2^{k_2} \notag\\
  &= \sum_{k_1,k_2=0}^\infty \frac{z_1 - z_2}{(k_1!)^2
    (k_2!)^2} \frac{1}{k_1 + k_2+1} z_1^{k_1} z_2^{k_2}\:.
\end{align}
Unfortunately, we have been unable to find an analytical proof for
arbitrary $N$ and $m$.  Partial analytical arguments are presented in
App.~\ref{sec:partialproof}, and a proof for $N=3$ is given in App.~\ref{sec:lsexpl},
but the most compelling evidence in favor
of the conjecture is due to numerical checks.  Using complex random
numbers for the $z_i$, we have checked the conjecture up to $N=25$ for
all values of $m$ with $1\le m\le N$.  We employed the GNU multi-precision
package \cite{gmp} and found the conjecture to be satisfied to a
precision of 2048 bits.  In the absence of a complete analytical
proof, this leads us to believe very strongly that the conjecture is
indeed correct.

A connection of Conjecture~\ref{th:ls} to Richardson-Littlewood
coefficients is discussed in App.~\ref{sec:rl}.

\subsection{Proof for a subset of coefficients in the case of
  arbitrary $N$ and $m$} 
\label{sec:partialproof} 

In this section we give a proof of the conjecture for a subset of
coefficients in the power series.  We note that $\J_0$ is
antisymmetric under the exchange of two of the $z_i$. Therefore, it is
divisible by the Vandermonde determinant $\Delta(z_1,\ldots,z_N)$,
which can be written as
\begin{equation}
  \Delta(z_1,\ldots,z_N) = \Delta(z_1,\ldots,z_m)
  \Delta(z_{m+1},\ldots,z_N) \prod_{i=1}^m \prod_{j=m+1}^N (z_i-z_j)\:.
\end{equation}
Hence $\J_0$ is also divisible by $\prod_{i=1}^m \prod_{j=m+1}^N
(z_i-z_j)$. The ansatz
\begin{equation}
  \label{eq:lspansatz}
  \J_0 = \left(\prod_{i=1}^m \prod_{j=m+1}^N (z_i-z_j)\right)
  \sum_{k_1,\ldots,k_N=0}^\infty c_{k_1,\ldots,k_N} z_1^{k_1} \cdots
  z_N^{k_N} \equiv \tilde{\J}_m
\end{equation}
yields
\begin{equation}
  c_{k_1,\ldots,k_N} \equiv \Delta(k_1,\ldots,k_m)
  \Delta(k_{m+1},\ldots,k_N) s_{k_1,\ldots,k_N}
\end{equation}
with $s_{k_1,\ldots,k_N}$ symmetric under the exchange of two of the
$k_i$ in the subsets $\lbrace k_1,\ldots,k_m \rbrace$ or $\lbrace
k_{m+1},\ldots,k_N \rbrace$ in order to satisfy the symmetry
requirements of $\J_0$ under exchange of the $z_i$.

In the following, we determine $s_{k_1,\ldots,k_N}$ in the special limit
\begin{equation}
0<z_1\ll z_2\ll \ldots \ll z_{m-1} \ll z_{m+1} \ll z_{m+2} \ll
\ldots \ll z_{N-1} \ll 1
\end{equation}
and $z_m$ and $z_N$ of the order of 1. In this limit the exchange term
simplifies to
\begin{equation}
  \prod_{i=1}^m \prod_{j=m+1}^N (z_i-z_j) \rightarrow
  (-1)^{(m-1)(N-m)} (z_m - z_N) z_m^{k_N^0} \prod_{j=m+1}^N
  z_j^{k_m^0}
\end{equation}
with
\begin{equation} 
  \begin{split}
    k_N^0 & \equiv N-m-1\:, \\
    k_m^0 & \equiv m-1\:.
  \end{split}
\end{equation}
The dominating terms in the power series in Eqs.~\eqref{eq:lspdef} and
\eqref{eq:lspansatz} correspond to
\begin{alignat}{3}
  \label{eq:lpslimit}
  k_i &= i-1 &\qquad& \text{for }\; i<m\:, \notag\\
  k_j &= j-m-1 && \text{for }\; m<j<N\:, \notag\\
  k'_i &= i-1 && \text{for }\; i<m\:, \notag\\ 
  k'_j &= j-2 && \text{for }\; m<j<N
\end{alignat}
with arbitrary $k_m, k_N, k'_m, k'_N$.  We therefore find
\begin{align}
  \tilde{\J}_m  &= (-1)^{(m-1)(N-m)} (z_m - z_N)
  \sum_{k_m=k_m^0,k_N=k_N^0}^\infty \Delta(k_1,\ldots,k_m)
  \Delta(k_{m+1},\ldots,k_N) s_{k_1,\ldots,k_N}  \notag\\
  &\quad \times \left(\prod_{i=1}^{k_m^0} z_i^{k_i} \right) \left(
    \prod_{j=m+1}^{N-1} z_j^{k_m^0+k_j} \right) z_m^{k_m+k_N^0}
  z_N^{k_N+k_m^0}\:, \\
  \J_0 &= \sum_{k'_m,k'_N=0}^\infty
  \frac{\Delta(k'_1,\ldots,k'_N)}{(k'_1!)^2 \cdots (k'_N!)^2} z_m^{k'_m}
  z_N^{k'_N} \left(\prod_{i=1}^{k_m^0} z_i^{k_i} \right) \left(
    \prod_{j=m+1}^{N-1} z_j^{k_m^0+k_j} \right)\:.
\end{align}
After inserting the fixed values of $k_i$, $k'_i$, $k_j$, and $k'_j$
for $i<m$ and $m<j<N$, we obtain that $\J_0 = \tilde{\J}_m$ is
equivalent to
\begin{align}
  \label{eq:lspstep2}
  \sum_{k'_m,k'_N=0}^\infty & \frac{(k'_N-k'_m)}{\Sigma_0 k'_m!
    (k'_m-(k_m^0+k_N^0))! k'_N! (k'_N-(k_m^0+k_N^0))!} z_m^{k'_m}
  z_N^{k'_N} \notag\\
  &=(z_N-z_m)\sum_{k_m=k_m^0,k_N=k_N^0}^\infty \frac{k_m! k_N!
    s_{0,1,\ldots,k_m^0-1,k_m,0,1,\ldots,k_N^0-1,k_N}}{(k_m-k_m^0)!
    (k_N-k_N^0)!} z_m^{k_m+k_N^0}  z_N^{k_N+k_m^0}
\end{align}
with
\begin{equation}
  \Sigma_0 \equiv 1! \cdots (k_m^0+k_N^0-1)!\: 1! \cdots (k_m^0-1)!\:
  1!\cdots (k_N^0-1)!\:,
\end{equation}
where we used the fact that
\begin{equation}
  \Delta(1+c,2+c,\ldots,N+c) = \left(\prod_{i=1}^{N-1} i!\right)
  (-1)^{\frac{N(N-1)}{2}}
\end{equation}
with an arbitrary constant $c$. Note that
\begin{equation}
  \frac{k_N-k_m}{k_m! (k_m-a)! k_N! (k_N-a)!} =
  \frac{1}{(k_N+k_m-a)} \frac{k_N(k_N-a)-k_m(k_m-a)}{k_m! (k_m-a)!
    k_N! (k_N-a)!}\:.
\end{equation}
Hence the left-hand side of Eq.~\eqref{eq:lspstep2} can be written as
\begin{align} 
  \sum_{k'_m,k'_N=0}^\infty & \frac{z_m^{k'_m}
    z_N^{k'_N}}{(k'_N+k'_m-(k_m^0+k_N^0))}
  \frac{k'_N(k'_N-(k_m^0+k_N^0))-k'_m(k'_m-(k_m^0+k_N^0))}{k'_m!
    (k'_m-(k_m^0+k_N^0))! k'_N! (k'_N-(k_m^0+k_N^0))!} \notag\\
  &=\sum_{k'_m,k'_N=0}^\infty
  \frac{(z_N-z_k)}{(k'_N+k'_m-(k_m^0+k_N^0)+1)} \frac{z_m^{k'_m}
    z_N^{k'_N}}{k'_m! (k'_m-(k_m^0+k_N^0))! k'_N!
    (k'_N-(k_m^0+k_N^0))!}\:.
\end{align}
Comparing coefficients of powers of $z_m$ and $z_N$ in
Eq.~\eqref{eq:lspstep2} now yields
\begin{align} 
  \label{eq:lpsspecials}
  s_{0,1,\ldots,k_m^0-1,k_m,0,1,\ldots,k_N^0-1,k_N} =
  \frac{1}{(k_m!)^2 (k_N!)^2 (k_m+k_N+1)} 
  \frac{1}{ \Sigma_0 \left(\prod_{i=1}^{k_N^0} (k_m+i) \right)
    \left(\prod_{j=1}^{k_m^0} (k_N+j) \right)}\:.
\end{align}
For the specific choice of $k_1,\ldots,k_N$ stated in Eq.~\eqref{eq:lpslimit} the desired general result,
\begin{equation}
  s_{k_1,\ldots,k_N} = \frac{1}{(k_1!)^2 \cdots (k_N!)^2} \prod_{i=1}^m
  \prod_{j=m+1}^N \frac{1}{k_i+k_j+1} \:,
\end{equation}
is equivalent to Eq.~\eqref{eq:lspstep2}. This means that for an infinite subset of coefficients, defined by Eq.~\eqref{eq:lpslimit}, the identity holds.

\subsection{Conjecture~\ref{th:ls} and Richardson-Littlewood
  coefficients}
\label{sec:rl}

In this section we discuss a relation for Richardson-Littlewood
coefficients that can be obtained from Conjecture~\ref{th:ls}.
Because of the antisymmetry of $\J_0$ and $\J_m$ under the exchange of
two of the $z_i$, we conclude that both sides are divisible by
$\Delta(z_1,\ldots,z_N)$. Therefore we define
\begin{align}
  \J'_0 &\equiv \frac{\J_0}{\Delta(z_1,\ldots,z_N)} 
  = \sum_{k_1>k_2>\dots>k_N \geq 0} 
  f_k \frac{\dete{z_i^{k_j}}{N}}{\Delta(z_1,\ldots,z_N)}\:,\\
  \J'_m &\equiv \frac{\J_m}{\Delta(z_1,\ldots,z_N)} 
  = \sum_{\begin{array}{c}
      \scriptstyle k^a_1>k^a_2>\dots>k^a_m \geq 0\\
      \scriptstyle k^b_{m+1}>k^b_{m+2}>\dots>k^b_N \geq 0
    \end{array}}
  g_{k^a k^b} \frac{\dete{z_i^{k^a_j}}{m}}{\Delta(z_1,\ldots,z_m)}
  \frac{\dete{z_{m+i}^{k^b_j}}{n}}{\Delta(z_{m+1},\ldots,z_N)}\:,
\end{align}
where we used Theorem~\ref{th:rearrangement}, $n\equiv N-m$ and
defined
\begin{align}
  f_k & \equiv  \frac{\Delta(k_1,\ldots,k_N)}{k_1!^2 \cdots k_N!^2}\:,\\
  g_{k^a k^b} & \equiv \frac{\Delta(k^a_1,\ldots,k^a_m)
    \Delta(k^b_1,\ldots,k^b_n)}{k^a_1!^2 \cdots k^a_m!^2 k^b_1!^2
    \cdots k^b_n!^2} \left( \prod_{i=1}^m \prod_{j=1}^n
    \frac{1}{k^a_i+k^b_j+1} \right)\:.
\end{align}
With $k^a_i \equiv p_i+m-i$, $k^b_i \equiv q_i+n-i$, and $k_i \equiv
r_i+N-i$ we can relate the $k_i$ to Young diagrams $p$, $q$, and
$r$. Note that
\begin{equation}
  \frac{\dete{z_i^{k_j}}{N}}{\Delta(z_1,\ldots,z_N)} \equiv
  S_r(z_1,\ldots,z_N)
\end{equation}
is the Schur function corresponding to partition $r$ \cite{MAC}.
Hence $\J_0 = \J_m$ resp. $\J'_0 = \J'_m$ can be written as
\begin{align} 
  \label{eq:rl}
  \sum_r f_r S_r(z_1,\ldots,z_N) = \sum_{p,q} g_{pq}
  S_p(z_1,\ldots,z_m) S_q(z_{m+1},\ldots,z_N)\:.
\end{align}
We now make use of the fact that Schur functions form a basis of the
ring of symmetric functions, i.e., the Schur functions on the
left-hand side can be expanded in Schur functions occurring on the
right-hand side. The expansion coefficients are called
Richardson-Littlewood coefficients $c^r_{\mu \nu}$ \cite{MAC},
\begin{equation}
  S_r(z_1,\ldots,z_N) = {\sum_{\makebox[0mm]{\scriptsize$\mu \subseteq r, \nu$}}}' c^r_{\mu \nu}
  S_\mu(z_1,\ldots,z_m) S_\nu(z_{m+1},\ldots,z_N)\:,
\end{equation}
where the sum is over all Young diagrams $\mu$ and $\nu$ with
\begin{equation}  
  \begin{split}
    \abs{\mu} + \abs{\nu} & = \abs{r} \:,\\
    \mu_i &\leq r_i \text{ for all } i\:.
  \end{split}
\end{equation}
As Schur functions are linearly independent, we can compare
coefficients in Eq.~\eqref{eq:rl} and find
\begin{equation} 
  \label{eq:rlrec}
  {\sum_{r \supseteq p}}' f_r c^r_{pq} = g_{pq}\:,
\end{equation}
where the sum is over all Young diagrams $r$ with $r_i \geq p_i $ for
all $i$ and $\abs{r} = \abs{p}+\abs{q}$. Thus Conjecture~\ref{th:ls}
implies the relation given in Eq.~\eqref{eq:rlrec} for
Richardson-Littlewood coefficients.

\section{The supersymmetric Leutwyler-Smilga integral for U$(1|1)$ and U$(2|1)$}
\label{sec:lsexpl}
In this section we calculate the supersymmetric Leutwyler-Smilga
integral, defined in Eq.~\eqref{eq:ILS}, over U$(1|1)$ and U$(2|1)$ by
explicit parametrization of the supergroup. This calculation is not
only a non-trivial check of our general result, but also serves as a proof of the power series identity which is conjectured in
App.~\ref{sec:dps} for $N=2$ and $N=3$.

The Leutwyler-Smilga integral with $\beta=1/2$ reads
\begin{equation}
\I \equiv \int d\mu(U) e^{\frac{1}{2} \str(A U+B U^\dagger)}\:,
\end{equation}
where the integral is over U$(m|n)$.  To prove Conjecture~\ref{th:ls}
it suffices to take $A$ and $B$ to be arbitrary diagonal supermatrices
with entries $a_1,\ldots,a_{m+n}$ and $b_1,\ldots,b_{m+n}$.

We adopt the following convention
for complex conjugation of anticommuting numbers,
\begin{align}
(\chi_1 \chi_2)^* & = \chi_2^* \chi_1^*\:,\\
(\chi_1^*)^* & = \chi_1\:,
\end{align}
and parametrize an element $U$ of the unitary supergroup U$(m|n)$ as
\begin{align}
 U = U_o ~ U_g = \left( \begin{array}{cc} U_m & 0 \\ 0 & U_n \end{array} \right)
\exp 
\left( 
\begin{array}{cccccc} 0 & \cdots & 0 & i \alpha_{11} & \cdots & i \alpha_{1n} \\
\vdots &  & \vdots & \vdots &  & \vdots \\
0 & \cdots & 0 & i \alpha_{m1} & \cdots & i \alpha_{mn} \\
i \alpha^*_{11} & \cdots & i \alpha^*_{m1} & 0 & \cdots & 0 \\
\vdots &  & \vdots & \vdots &  & \vdots \\
i \alpha^*_{1n} & \cdots & i \alpha^*_{mn} & 0 & \cdots & 0
 \end{array} \right),
\end{align}
where $U_m$ and $U_n$ are ordinary unitary $m \times m$ and $n \times n$ matrices and $\alpha_{11},\ldots,\alpha_{mn}$ are anticommuting variables.
This generalizes the parametrization introduced in
Refs.~\cite{Guhr:1992ih, Guhr:1992ei} for U$(1|1)$. It is also very
similar to the parametrization of the super-Riemannian manifold Gl$(m|1)$ used in Ref.~\cite{Damgaard:1998xy}.

The invariant integration measure corresponding to this
parametrization for the cases of $m=n=1$ and $m=2$, $n=1$ is of the form
\begin{align}
	d\mu(U) = d\mu(U_m) ~ d\mu(U_m) ~ d\alpha_{11} d\alpha^*_{11} \cdots d\alpha_{mn} d\alpha^*_{mn} ~\mathcal{T}_{m,n}(\alpha_{11},\ldots,\alpha_{mn},\alpha_{11}^*,\ldots,\alpha_{mn}^*)\:,
\end{align}
where $d\mu(U_m)$ and $d\mu(U_n)$ are the invariant measures of the ordinary groups U$(m)$ and U$(n)$, and $\mathcal{T}_{m,n}$ is a function only of anticommuting variables.
We find
\begin{align}
\mathcal{T}_{1,1}(\alpha_{11},\alpha_{11}^*) & = 1\:, \\
\mathcal{T}_{2,1}(\alpha_{11},\alpha_{21},\alpha_{11}^*,\alpha_{21}^*) & = 1-\frac{1}{3} \left(\alpha_{11} \alpha_{11}^* + \alpha_{21} \alpha_{21}^*\right)\:.
\end{align}
The supertrace can be written as
\begin{align}\label{eq:lsexponentidentity}
	\str(A U+B U^\dagger) = \str(U_g A U_o+B U_g^\dagger U_o^\dagger) = \tr(\tilde{A}_m U_m+\tilde{B}_m U_m^\dagger) - \tr(\tilde{A}_n U_n+\tilde{B}_n U_n^\dagger)\:,
\end{align}
where $\tilde{A}_m$ and $\tilde{B}_m$ are the boson-boson blocks of the supermatrices $U_g A$ and $B U_g^\dagger$, and $\tilde{A}_n$ and $\tilde{B}_n$ are the fermion-fermion blocks of the supermatrices $U_g A$ and $B U_g^\dagger$.
The integral thus factorizes in the following way,
\begin{align} \nonumber
\I = & \int d\alpha_{11} d\alpha^*_{11} \cdots d\alpha_{mn} d\alpha^*_{mn} \mathcal{T}_{m,n}(\alpha_{11},\ldots,\alpha_{mn},\alpha_{11}^*,\ldots,\alpha_{mn}^*) \\ 
 	&\times \left(\int d\mu(U_m) e^{\frac{1}{2} \tr(\tilde{A}_m U_m+\tilde{B}_m U_m^\dagger)} \right) \left(\int d\mu(U_m) e^{ -\frac{1}{2} \tr(\tilde{A}_n U_n+\tilde{B}_n U_n^\dagger)} \right).
\end{align}
We perform the ordinary Leutwyler-Smilga integrals over the groups U$(m)$ and U$(n)$ \cite{SWa} and are left with
\begin{align} \nonumber
\I = & \int d\alpha_{11} d\alpha^*_{11} \cdots d\alpha_{mn} d\alpha^*_{mn} \mathcal{T}_{m,n}(\alpha_{11},\ldots,\alpha_{mn},\alpha_{11}^*,\ldots,\alpha_{mn}^*) \\
 	&\times \left( \frac{\dete{\lambda_j^{m-i} I_{m-i}(\lambda_j)}{m}}
	  {\Delta(\lambda^2_1,\ldots,\lambda^2_m)} \right) \left( \frac{\dete{\mu_j^{n-i} I_{n-i}(\mu_j)}{n}}
	  {\Delta(\mu^2_1,\ldots,\mu^2_n)}\right),
\end{align}
where $\lambda^2_1,\ldots,\lambda^2_m$ are the eigenvalues of the
matrix $\tilde{A}_m \tilde{B}_m$, $\mu^2_1,\ldots,\mu^2_n$ are the eigenvalues of the
matrix $\tilde{A}_n \tilde{B}_n$, and $I_\nu$ is the modified Bessel function of the
first kind. Furthermore, $\Delta(\lambda^2_1,\ldots,\lambda^2_m) \equiv \prod_{1\leq i<j\leq m}(\lambda^2_i-\lambda^2_j)$ is the Vandermonde determinant.

In the case of U$(1|1)$ this is equal to
\begin{align}
\I = \int d\alpha_{11}^* d\alpha_{11} I_0([a_1 b_1 (1 + \alpha_{11}^* \alpha_{11})]^{1/2}) I_0([a_2 b_2 (1 - \alpha_{11}^* \alpha_{11})]^{1/2})\:.
\end{align}
We expand the Bessel functions in order to obtain the linear term in $\alpha_{11}^*\alpha_{11}$,
\begin{align}
I_0([a_1 b_1 (1 + \alpha_{11}^* \alpha_{11})]^{1/2}) & = \sum_{k=0}^\infty \frac{1}{(k!)^2} \left(\frac{a_1 b_1}{4}\right)^k \left(1 + k \alpha_{11}^* \alpha_{11}\right)\:,\\
I_0([a_2 b_2 (1 - \alpha_{11}^* \alpha_{11})]^{1/2}) & = \sum_{\ell=0}^\infty \frac{1}{(\ell!)^2} \left(\frac{a_2 b_2}{4}\right)^\ell \left(1 - \ell \alpha_{11}^* \alpha_{11}\right)\:.
\end{align}
Thus, we can write
\begin{align}
\I & = \frac{1}{2}  \left( \sqrt{a_1 b_1} \frac{\partial}{\partial \sqrt{a_1 b_1}} - 
\sqrt{a_2 b_2} \frac{\partial}{\partial \sqrt{a_2 b_2}}\right) I_0(\sqrt{a_1 b_1}) I_0(\sqrt{a_2 b_2}) \\
& = \frac{1}{2} \det \left( \begin{array}{cc} I_0(\sqrt{a_1 b_1}) & \sqrt{a_1 b_1} I_1(\sqrt{a_1 b_1}) \\ I_0(\sqrt{a_2 b_2}) & \sqrt{a_2 b_2} I_1(\sqrt{a_2 b_2})  \end{array} \right).
\end{align}
This is equivalent to the special case of our general result given in
Eq.~\eqref{eq:ls_result}. 

Let us now turn to the case of U$(2|1)$, which is slightly more involved.
In this case the calculation of the eigenvalues yields
\begin{align}
	\lambda_1^2 & = a_1 b_1 \left(1-\alpha_{11}\alpha^*_{11}-\frac{a_1 b_1 + 2 a_2 b_2}{3 (a_1 b_1-a_2 b_2)} \alpha_{11}\alpha^*_{11} \alpha_{21}\alpha^*_{21} \right), \\
	\lambda_2^2 & = a_2 b_2 \left(1-\alpha_{21}\alpha^*_{21}+\frac{2 a_1 b_1 + a_2 b_2}{3 (a_1 b_1-a_2 b_2)} \alpha_{11}\alpha^*_{11}\alpha_{21}\alpha^*_{21} \right), \\
	\mu_1^2 & = a_3 b_3 \left(1+\alpha_{11}\alpha^*_{11}+\alpha_{21}\alpha^*_{21}+\frac{2}{3} \alpha_{11}\alpha^*_{11} \alpha_{21}\alpha^*_{21} \right).
\end{align}
The calculation of the determinants of Bessel functions of these eigenvalues is tedious but straightforward. Finally, we find that the term proportional to $\alpha_{11}\alpha^*_{11} \alpha_{21}\alpha^*_{21}$ is indeed
\begin{align}
\frac{1}{a_1 b_1 - a_2 b_2} \det \left( \begin{array}{ccc} 
		I_0(\sqrt{a_1 b_1}) & I_0(\sqrt{a_2 b_2}) & I_0(\sqrt{a_3 b_3}) \\
		\sqrt{a_1 b_1} I_1(\sqrt{a_1 b_1}) & \sqrt{a_2 b_2} I_1(\sqrt{a_2 b_2}) & \sqrt{a_3 b_3} I_1(\sqrt{a_3 b_3}) \\
		a_1 b_1 I_2(\sqrt{a_1 b_1}) & a_2 b_2 I_2(\sqrt{a_2 b_2}) & a_3 b_3 I_2(\sqrt{a_3 b_3})
	\end{array}\right).
\end{align}
This is again equivalent to the special case of our general result given in Eq.~\eqref{eq:ls_result}. Note that this result also proves the
power series identity conjectured in App.~\ref{sec:dps} for $m=2$,
$n=1$. We do not present the calculation for $m=1$, $n=2$ here since
it is very similar to the case of $m=2$, $n=1$. The next step would be to perform this procedure for general U($m|n$), which would
then also yield a proof of the conjecture for arbitrary values of $m$ and $n$.

\end{document}